\documentclass[twocolumn,showpacs,preprintnumbers,amsmath,amssymb,superscriptaddress]{revtex4-1}
\usepackage{graphicx}
\usepackage{subfigure}
\usepackage{bm}

\usepackage[usenames]{color}

\begin{document}


\title{
Theory of inplane magnetoresistance in two-dimensional massless Dirac fermion system
}


\author{Takao Morinari}\email{morinari@yukawa.kyoto-u.ac.jp}
\author{Takami Tohyama}
\affiliation{Yukawa Institute for Theoretical Physics, Kyoto University, Kyoto 606-8502, Japan}

\date{\today}

\begin{abstract}
We present the theory of the inplane magnetoresistance
in two-dimensional massless Dirac fermion systems
including the Zeeman splitting and the electron-electron interaction effect 
on the Landau level broadening within a random phase approximation.
With the decrease in temperature, 
we find a characteristic temperature dependence of the inplane magnetoresistance 
showing a minimum followed by an enhancement with a plateau.
The theory is in good agreement with the experiment of the layered 
organic conductor $\alpha$-(BEDT-TTF)$_2$I$_3$ under pressure.
In-plane magnetoresistsnce of graphene is also discussed based on this theory.
\end{abstract}

\pacs{
73.43.Qt, 
71.10.Pm  
71.70.-d, 
72.15.Gd  
}

\maketitle

\newcommand{\be}{\begin{equation}}
\newcommand{\ee}{\end{equation}}
\newcommand{\bea}{\begin{eqnarray}}
\newcommand{\eea}{\end{eqnarray}}

\section{Introduction}

Since the discovery of unconventional integer quantum Hall 
effect in graphene,\cite{Novoselov2005,Zhang2005}
which is a single atomic sheet of graphite,
massless Dirac fermions realized in condensed matter systems have attracted much attention.
Under magnetic field, a remarkable difference between conventional 
two-dimensional electron systems and two-dimensional Dirac fermion systems appears 
in the Landau level structure.
In conventional electrons, the Landau level energies are equally spaced.
Meanwhile the Landau level energies in Dirac fermions with the Fermi velocity $v$
are given by
\be
E_n = {\rm sgn}(n) \frac{\hbar v}{\ell_B}{\sqrt{2|n|}},
\label{eq_LL}
\ee
where $n=0, \pm 1, \pm 2, ...$ and $\ell_B = \sqrt{\hbar/eB}$ 
is the magnetic length.\cite{CastroNeto09}
For the case of Dirac fermions, the Landau levels are unevenly spaced.
What makes a crucial difference compared to the case of conventional electrons
is the existence of the zero energy Landau level
that plays a central role for the unconventional 
integer quantum Hall effect.\cite{Geim07}

Massless Dirac fermion systems are not restricted to a purely two-dimensional system.
The layered organic conductor $\alpha$-(BEDT-TTF)$_2$I$_3$ under pressure
shows remarkable physical properties associated 
with a Dirac fermion spectrum.\cite{Tajima06}
Theoretically it has been predicted that this system is a massless Dirac fermion system \cite{Kobayashi04,Katayama06} 
where the Fermi energy is at the Dirac point
and the Dirac cone is tilted.\cite{Kobayashi07,Goerbig08,Morinari09}
This massless Dirac fermion spectrum is supported by
first principles calculations.\cite{Ishibashi2006,Kino2006}
Experimentally the observation of the negative interlayer 
magnetoresistance \cite{Tajima09} supports
the massless Dirac fermion spectrum.
Application of the magnetic field decreases the interlayer resistivity.
This negative interlayer magnetoresistance is consistent with
the existence of the zero energy Landau level.\cite{Osada08}
The interlayer resistance decreases in proportion to the inverse 
of the applied magnetic field.
This magnetic field dependence arises from 
the zero energy Landau level degeneracy.

In this organic Dirac fermion system, an intriguing inplane 
magnetoresistance was observed.\cite{Tajima06}
Under magnetic field, 
the inplane resistivity decreases gradually
as the temperature $T$ is decreased for $T>100$K.
After reaching a broad minimum around $100$K, the resistivity increases
and then shows a narrow plateau region around several Kelvin.
After that the resistivity increases again as the temperature is decreased further.

In this paper, we present the theory of the inplane magnetoresistance 
in massless Dirac fermion systems
including the Landau level broadening effect
due to the Coulomb interaction between Dirac fermions
and the Zeeman energy splitting.
We compute the inplane longitudinal conductivity by the Kubo formula using the 
Landau level wave functions for massless Dirac fermions.
The Coulomb interaction effect on the Landau level broadening
is computed by the random phase approximation.
The result is consistent with the inplane magnetoresistance
observed in $\alpha$-(BEDT-TTF)$_2$I$_3$.\cite{Tajima06}
The theory is also applied to graphene.

\section{Model}
\label{sec_model}
For the description of two-dimensional Dirac fermions in the $x-y$ plane,
we introduce two component spinor field operator $\psi_{\sigma} (x,y)$ 
where ${\sigma} = \pm$ denotes the spin.
In graphene and $\alpha$-(BEDT-TTF)$_2$I$_3$,
there are two Dirac points in the Brillouin zone.
We assume that Dirac fermions are degenerate with respect to these valley
degrees of freedom.
We do not consider inter-valley interaction and
focus on the single-valley properties.
The Hamiltonian is given by $\mathcal{H} = \mathcal{H}_0 + \mathcal{V}_C$,
where
\be
\mathcal{H}_0 = \sum_{\sigma} \int {dx} \int {dy} \,\psi_{\sigma} ^\dag  
\left( {x,y} \right)\hbar v\left( {\widehat{k}_x \sigma _x  + \widehat{k}_y \sigma _y } \right)
\psi_{\sigma} \left( {x,y} \right),
\ee
with $\widehat{k}_{x,y} = -i\partial_{x,y}$ and $\sigma_{x,y}$ the Pauli matrices.
The term $\mathcal{V}_C$ describes 
the Coulomb interaction between Dirac fermions,
$\mathcal{V}_C  = 
(1/2) \sum\limits_{\mathbf{q}} {V_{\mathbf{q}}}
\rho _{\mathbf{q}} \rho _{ - {\mathbf{q}}}$,
where $V_q = 2\pi e^2/(4\pi \epsilon_0 \epsilon |\mathbf{q}|)$
with $\epsilon$ the dielecric constant.
Throughout this paper, we assume that the Fermi energy is at the Dirac point.
We do not include the effect of the Dirac cone tilt in 
$\alpha$-(BEDT-TTF)$_2$I$_3$ (Ref. \cite{Kobayashi07}) 
because it turns out that tilt is unimportant for understanding 
the main features of the inplane magnetoresistance of
$\alpha$-(BEDT-TTF)$_2$I$_3$ as we shall see below.

In a magnetic field, the kinetic energy of Dirac fermions is quantized
into Landau levels, Eq.~(\ref{eq_LL}).
Taking the Landau gauge ${\bf A}=(0,Bx)$, 
the Landau level wave functions are represented by
$\Phi _{n,k} \left( {x,y} \right) = 
\exp \left( {iky} \right)\phi _{n,k} \left( x \right)
/\sqrt{L_y}$,
where $L_y$ is the system size in the $y$-direction and
\bea
\phi _{n,k} \left( x \right) &=& \frac{{C_n }}
{{\sqrt {\ell _B } }}\left[ \left( {\begin{array}{*{20}c}
   { - i\operatorname{sgn} n}  \\
   0  \\
 \end{array} } \right)h_{\left| n \right| - 1} \left( {\frac{x}
{{\ell _B }} + k\ell _B } \right) 
\right. \nonumber \\ 
& & \left. + \left( {\begin{array}{*{20}c}
   0  \\
   1  \\
 \end{array} } \right) h_{\left| n \right|} \left( {\frac{x}
{{\ell _B }} + k\ell _B } \right) \right],
\eea
with $C_0 = 1$ and $C_n = 1/\sqrt{2}$ for $n \neq 0$,
and ${\rm sgn}n=1 (-1)$ for $n>0 (n<0)$ and 
${\rm sgn}n=0$ for $n=0$.

Here $h_n (\xi)$ are the eigenstates of the harmonic oscillator Hamiltonian
$-\partial_{\xi}^2/2+{\xi}^2/2$,
$h_n \left( \xi  \right) = 
H_n \left( \xi  \right)\exp 
\left( { - \xi ^2 /2} \right)
/\left( {2^{n/2} \pi ^{1/4} \sqrt {n!}} \right)
,$
with $H_n(\xi)$ the Hermite polynomial.

In terms of the Landau level wave functions,
the field operator $\psi_{\sigma} (x,y)$ is represented by
$\psi_{\sigma} \left( {x,y} \right) 
= \sum\limits_{n,k} {\Phi _{n,k} \left( {x,y} \right)c_{n,k,\sigma} }$.
Using this form, we find that the Fourier transform of the density operator
$\rho (x,y) = \sum_{\sigma} \psi^{\dagger}_{\sigma}(x,y) \psi_{\sigma} (x,y)$ is
\bea
\rho _{\mathbf{q}} &=& e^{ - \frac{{q^2 \ell _B^2 }}
{4}} e^{\frac{i}
{2}q_x q_y \ell _B^2 } \sum\limits_{n_1 ,n_2 ,k,\sigma} e^{iq_x k\ell _B^2 } \nonumber \\
& & \times F_{n_1 ,n_2 } \left( {\bf q} \right)c_{n_1,k,\sigma}^\dag  c_{n_2 ,k + q_y,\sigma} ,
\eea
where the function $F_{n_1 ,n_2 }({\bf q})$ 
is defined by\cite{Shizuya07,*Roldan09,*Shizuya10}
\bea
F_{n_1 ,n_2} \left( {\bf q} \right) &=& 
C_{n_1 } C_{n_2 } \left[ J_{\left| {n_1 } \right|,\left| {n_2 } \right|} 
\left( {\bf q} \right) \right. \nonumber \\
& & \left. + \operatorname{sgn} \left( {n_1 n_2 } \right)
J_{\left| {n_1 } \right| - 1,\left| {n_2 } \right| - 1} \left( {\bf q} \right) \right].
\eea
For $n_1>n_2$, the function $J_{n_1 ,n_2 } \left( {\bf q} \right)$ 
has the following form
\bea
J_{n_1 ,n_2 } \left( {\bf q} \right) &=& \sqrt {\frac{{n_1!}}
{{n_2 !}}} \left( {\frac{{ - iq_x  - q_y }}
{{\sqrt 2 }}\ell _B } \right)^{n_1   - n_2  } \nonumber \\
& & \times L_{n_2 }^{n_1 - n_2} \left( {\frac{{q^2 \ell _B^2 }}
{2}} \right),
\eea
and $J_{n_2,n_1}({\bf q})=\left[ J_{n_1,n_2}(-{\bf q}) \right]^*$.
Here $L_n^m(x)$ are the associated Laguerre polynomials.

\section{The Coulomb interaction effect on the Landau level broadening}
Now we compute the Coulomb interaction effect on the scattering rate of Dirac fermions
that leads to the Landau level broadening.
As we shall show below the temperature dependence of the Landau level broadening
gives rise to a broad minimum in the inplane 
resistivity that appears around $T=T_{\rm min}$.
(For the case of $\alpha$-(BEDT-TTF)$_2$I$_3$, 
it has been reported\cite{Tajima06} that
$T_{\rm min} \sim 100$K.)
Although it is easy to include the Zeeman splitting
in the calculation of the Landau level broadening,
we present the calculation for the spinless case
because the interaction effect plays an important role
at high temperatures where many Dirac fermions are excited from 
the zero energy Landau level while
the Zeeman spin splitting effect is negligible.

The single particle Matsubara Green's function for the Landau level with the index $n$
is 
$G_{n} \left( i \omega_{\nu} \right) =  
1/\left[ i \omega_{\nu} - E_{n} - \Sigma_n ( i \omega_{\nu} ) \right]$,
where $\omega_{\nu} = (2\nu + 1)\pi k_B T$ is the fermion Matsubara frequency.
Within the random phase approximation, the self-energy $\Sigma_n (i \omega_{\nu} )$
is described by
\bea
\Sigma _n \left( {i\omega _\nu  } \right) &=&  - \frac{k_B T}
{{2\pi \ell _B^2 }}\sum\limits_{{\mathbf{q}},n',i\Omega _{\nu '} } 
{\frac{{V_{\mathbf{q}} }}
{{1 - V_{\mathbf{q}} D_{\mathbf{q}} \left( {i\Omega _{\nu '} } \right)}}} 
\nonumber \\
& & \times F_{n,n'} \left( {\mathbf{q}} \right)F_{n',n} \left( { - {\mathbf{q}}} \right)
G_{n'} \left( {i\omega _\nu   + i\Omega _{\nu '} } \right),
\label{eq_Sigma}
\eea
where
\bea
D_{\mathbf{q}} \left( {i\Omega _\nu  } \right) &=&  - \frac{{e^{ - \frac{{q^2 \ell _B^2 }}
{2}} }}
{{2\pi \ell _B^2 }}\sum\limits_{n_1 ,n_2 } F_{n_1 ,n_2 } 
\left( {\mathbf{q}} \right)F_{n_2 ,n_1 } \left( { - {\mathbf{q}}} \right)
\nonumber \\ 
& & \times \frac{{f\left( {E_{n_1 } } \right) - f\left( {E_{n_2 } } \right)}}
{{i\Omega _\nu   - E_{n_1 }  + E_{n_2 } }}.
\eea
The summation over the boson Matsubara frequency $\Omega_{\nu'}$
in Eq.~(\ref{eq_Sigma}) is carried out by using the spectral representation
of $V_{\bf q} (i\Omega_{\nu}) \equiv 
V_{\bf q}/\left[1-V_{\bf q} D_{\bf q}(i\Omega_{\nu}) \right]$.
Performing the analytic continuation $i \omega_{\nu} \rightarrow \omega + i\delta$
with $\delta$ an infinitesimal number and after some algebra, we obtain
\bea
\Sigma _n \left( {\omega  + i\delta } \right) 
&=&  - \frac{1}
{{4\pi ^3 }\ell_B^2 }\int_{ - \infty }^\infty  {d\varepsilon } \int_0^\infty  {dq} q
\operatorname{Im} \left[ V_{\bf q}({\varepsilon  + i\delta })
\right] \nonumber \\
& & \hspace{-5em} \times \sum\limits_{n'} {F_{n,n'} \left( {\mathbf{q}} \right)F_{n',n} 
\left( { - {\mathbf{q}}} \right)
\frac{{n\left( \varepsilon  \right) + f\left( {E_{n'} } \right)}}
{{\omega  + i\delta  - E_{n'}  + \varepsilon }}}.
\eea
The imaginary part of the self-energy,
$- \operatorname{Im} \Sigma _n \left( {\omega  + i\delta } \right)$,
leads to the Landau level broadening.
Instead, we use an approximate form,
$\Gamma_n^C \equiv - \operatorname{Im} \Sigma _n 
\left( {E_n  + i\delta } \right)$.
We do not attempt to compute this quantity in a self-consistent manner.
The Coulomb interaction plays an important role if there are 
large numbers of excited Dirac fermions.
However, the number of the excited Dirac fermions
is suppressed at temperatures less than the Landau level energy gap.
In such a regime, we may treat the Coulomb interaction perturbatively.

\begin{figure}[htbp]
  \begin{center}
  \subfigure[]{
    \includegraphics[width=0.8 \linewidth]{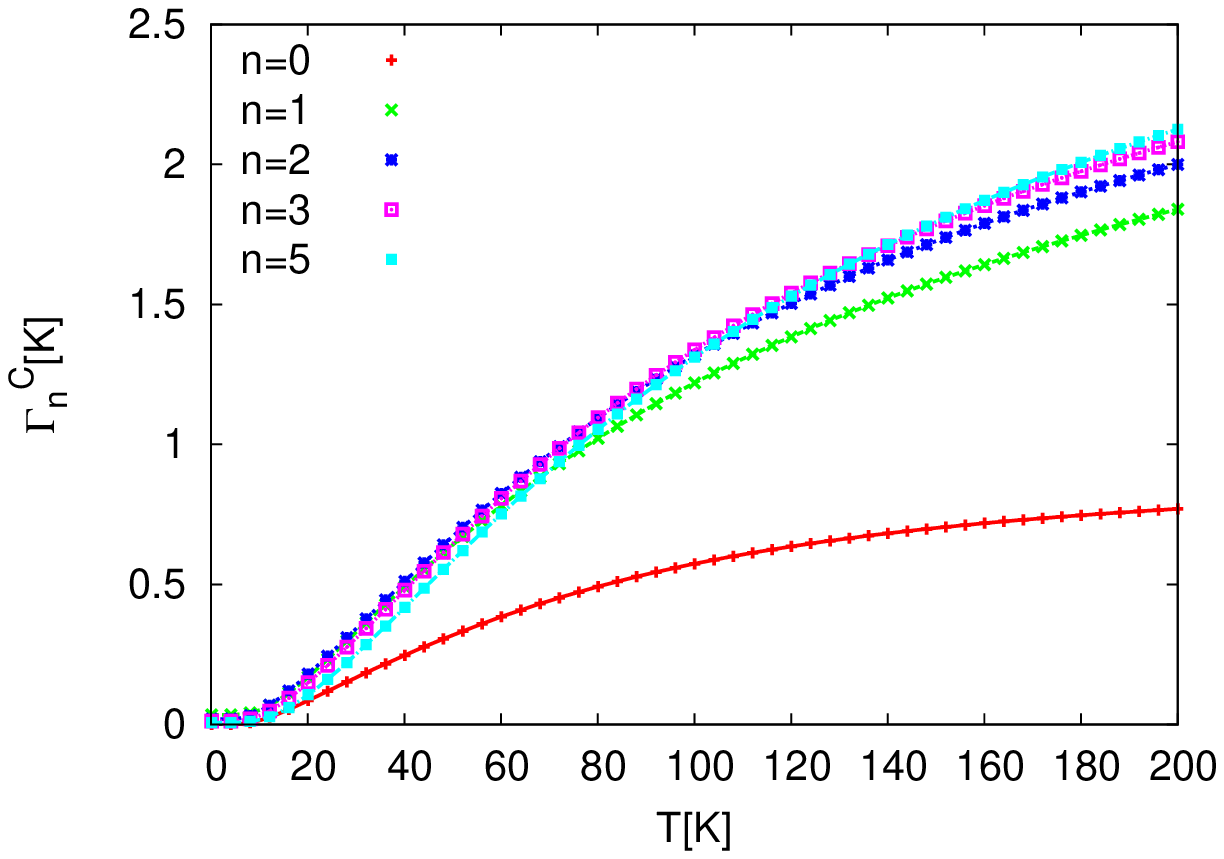}
  }
  %
  \subfigure[]{
    \includegraphics[width=0.8 \linewidth]{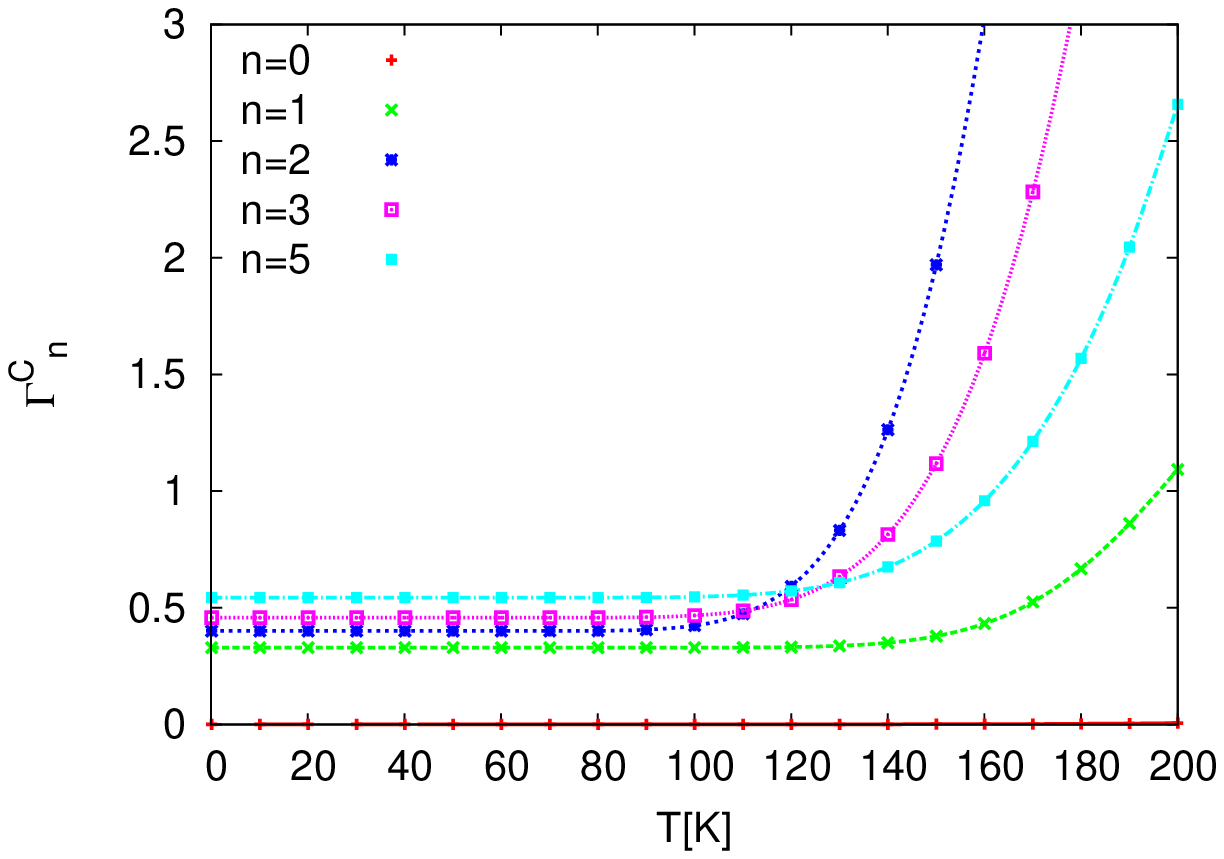}
  }
\end{center}
   \caption{ \label{fig_scattering_rate}
	(Color online)
	(a) Temperature dependence of $\Gamma_n^C$ at $B=10$T
        for different Landau levels with $\delta=0.1$
	for $\alpha$-(BEDT-TTF)$_2$I$_3$.
	(b)Temperature dependence of $\Gamma_n^C$ at $B=10$T
	for graphene.
    }
\end{figure}

Figure \ref{fig_scattering_rate} (a) shows $\Gamma_n^C$ for
different Landau levels where we set \cite{Morinari10a} 
$\sqrt{2/B}\hbar v/\ell_B = 10$K/T$^{-1/2}$
and $\epsilon = 300$ that were estimated \cite{Tajima10prb} 
from the analysis of the interlayer magnetoresistance 
in $\alpha$-(BEDT-TTF)$_2$I$_3$.
(Note that at ambient pressure
a large dielectric constant that is the same order of magnitude
as our value has been reported.\cite{Ivek10} )

In the numerical calculation, we used the recursion formula for the 
function 
$\sqrt{ n!/(n+k)! }
x^{k/2} \exp \left( { - x/2} \right)L_n^k \left( x \right)$
instead of the recursion formula for the associated Laguerre polynomials
because $L_n^k(x)$ and the factorials can be huge for Landau levels with $|n| \gg 1$.
The summation with respect to the Landau
levels is taken from $n = -50$ to $50$.
At temperatures below $\sim 10$~K, $\Gamma_n^C$ remain constant.
This behavior is understood from the energy gaps created 
by the Landau level structure:
the Coulomb interaction plays an important role 
when there are excited Dirac fermions to higher Landau levels. 
In order to excite Dirac fermions to higher Landau levels,
the temperature should be larger than the energy gap created by
the Landau levels. 
Thus, the characteristic temperatures are determined 
from the energy gaps between the Landau levels.
As shown in Fig.~\ref{fig_scattering_rate}(a), 
$\Gamma_0^C$ behaves remarkably differently
while the other $\Gamma_n^C$ ($n \neq 0$) behaves similarly.
At low temperature below $30$~K the effect of the electron-electron interaction
is rapidly suppressed because of the large energy gap between the zero-energy Landau level
and the $|n|=1$ Landau level.
Reflecting this fact, $\Gamma_n^C$ decreases as we increase the magnetic field
because the Landau level energy gaps increase.

Figure \ref{fig_scattering_rate}(b) shows $\Gamma_n^C$ for graphene where
we take $\epsilon=2.5$ for the dielectric constant \cite{Giesbers09} 
and $\sqrt{2/B}\hbar v/\ell_B = 400$K/T$^{-1/2}$ for the Landau level
structure parameter.\cite{Geim07}
Although the temperature dependence of $\Gamma_n^C$ is different from
Fig.~\ref{fig_scattering_rate}(a) because of the parameter differences,
it is common that the $n=0$ Landau level component behaves differently
as compared with the n=0 Landau level.
Since the Landau level energy gaps between the $n=0$ Landau level and 
the $n=1$ Landau level for graphene is about $1000$~K,
the value of $\Gamma_0^C$ is negligible in the temperature range
shown in Fig.~\ref{fig_scattering_rate}(b). 

This result is consistent with the experiment \cite{Giesbers07} 
suggesting that the zero energy Landau level is quite sharp in shape compared 
with the other Landau levels.
Compared to $\alpha$-(BEDT-TTF)$_2$I$_3$, the almost temperature 
independent region extended until $\sim 120$~K.
This is because the Landau level energy spacing in graphene 
is larger than that in $\alpha$-(BEDT-TTF)$_2$I$_3$.
As shown in Fig.~\ref{fig_scattering_rate}(b) the interaction effect on $\Gamma_n^C$
is negligible for $T<100$ K due to the large separation 
between the Landau levels.

\section{Inplane magnetoresistance}
\label{sec_inplane_mr}
Now we compute the inplane longitudinal conductivity 
$\sigma_{xx}$ using the Kubo formula,\cite{Shon98}
\bea
\sigma _{xx}  &=& \frac{{e^2 }}
{\hbar }\left( {\frac{{\hbar v}}
{{\ell _B }}} \right)^2 \sum\limits_{n,\sigma} {C_n } 
\int_{ - \infty }^\infty  {dE} \left( { - \frac{{\partial f}}
{{\partial E}}} \right) \nonumber \\
& & \times 
\frac{{\Gamma _n /\pi }}{{\left( {E - E_{n,\sigma } } \right)^2  + \Gamma _n^2 }}
\left[ {\frac{{\Gamma _{\left| n \right| + 1} /\pi }}
{{\left( {E - E_{\left| n \right| + 1,\sigma } } \right)^2  
+ \Gamma _{\left| n \right| + 1}^2 }}} \right. \nonumber \\
& & \left. 
+ \frac{{\Gamma _{ - \left| n \right| - 1} /\pi }}
{{\left( {E - E_{ - \left| n \right| - 1,\sigma } } \right)^2  
+ \Gamma _{ - \left| n \right| - 1}^2 }} \right],
\label{eq_kubo_formula}
\eea
where $f$ is the Fermi distribution function and 
the Zeeman energy splitting is included as
$E_{n,\sigma }  = E_n  + g\mu _B \sigma B/2$.
Here $\mu_B$ is the Bohr magneton and 
we set $g=2$.
The scattering rate is assumed to be
$\Gamma_n = \Gamma_0 + \Gamma_n^C$,
where $\Gamma_0$ is associated with impurity scattering.
In the following calculation we take $\Gamma_0 = 2$K 
that was estimated from analysis of the interlayer 
magnetoresistance data \cite{Tajima09} 
at low temperatures.\cite{Morinari10a}
To reduce the numerical computation time we use P{\'{a}}de approximants 
for the temperature dependence of $\Gamma_n^C$.
For Landau levels with $n\neq 0$ we used the same P{\'{a}}de approximant for $\Gamma_1^C$
because $\Gamma_n^C$ with $n \neq 0$ behave similarly as shown 
in Fig.~\ref{fig_scattering_rate}(a).

\begin{figure}
   \begin{center}
    \includegraphics[width=0.8 \linewidth]{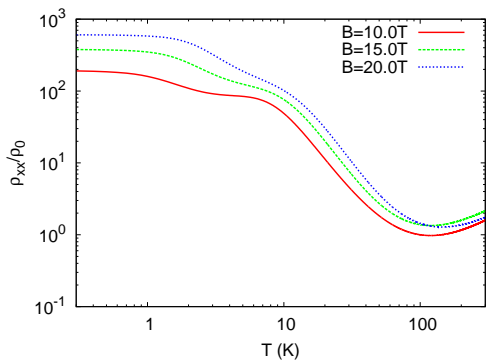}
   \end{center}
   \caption{ \label{fig_inplane_mr}
	(Color online)
	The inplane resistivity for different magnetic fields with $\Gamma_0=2$~K.
	The normalization parameter $\rho_0$ is taken as 
	$\rho_0 = \rho_{xx}(100{\rm K})$ at $B=10$T to 
	compare with the experiment in Ref.~\cite{Tajima06}.
    }
 \end{figure}

Figure \ref{fig_inplane_mr} shows the inplane resistivity, $\rho_{xx} = 1/\sigma_{xx}$
for different magnetic fields.
Note that $\sigma_{xy}=0$ because the Fermi energy is at the Dirac point.
Here we assume particle-hole symmetry so that the Fermi energy is fixed to the Dirac point
even at finite temperatures.
The minima appear around $T_{\rm min} \simeq 100$K.
These minima appear because of the onset of the Landau level splitting effect:
The Landau levels with $|n|<10$ are well separated each other.
But those separations are unimportant for $T \sim 100$~K
because of the temperature broadening effect due to the derivative of 
the Fermi distribution function in Eq.~(\ref{eq_kubo_formula}).
For $T>100$~K, Landau levels with $|n| \leq 10$ are almost continuously distributed
because $|E_{n+1}-E_n|< \Gamma_{n+1} + \Gamma_n$.
For $T<100$~K, we find that 
$|E_{10 \pm 1}-E_{10}| > \Gamma_{10 \pm 1} + \Gamma_{10}$
from the temperature dependence of $\Gamma_n^C$.
So the Landau level splitting effect appears for $T<100$K.
We computed $\sigma_{xx}$ without including $\Gamma_n^C$,
and confirmed that the temperature dependence of $\rho_{xx}$ 
for $T>100$K mainly arises from the temperature dependence of $\Gamma_n^C$.

The appearance of a minimum at a characteristic temperature $T_{\rm min}$
in the inplane magnetoresistance suggests that $T_{\rm min}$
is a crossover temperature from the interaction dominant regime to
the almost non-interacting regime:
for $T > T_{\rm min}$, the Landau level broadening
smears out the Landau level energy spectrum.
In this regime, the Landau level spacing is unimportant,
and the electron-electron interaction, which requires 
the excitations from one Landau level to higher Landau levels,
plays an important role.
By contrast for $T < T_{\rm min}$, the Landau level broadening
is less than the Landau level spacing.
Thus, the excitations from one Landau level to higher Landau levels
are suppressed.
The characteristic temperature $T_{\rm min}$ 
depends on $\epsilon$, $v$, and $B$.
Although there is no simple analytical formula for $T_{\rm min}$,
one can determine $T_{\rm min}$
from the inplane magnetoresistance measurement. 
The same analysis can be applied to the surface states of three dimensional 
topological insulators.\cite{FuKaneMele07,MooreBalents07}

With decreasing the temperature from $\sim 100$K the resistivity increases
because the number of Landau levels contributing to $\sigma_{xx}$ decreases.
Below $10$K a narrow plateau region appears.
If we compute $\rho_{xx}$ omitting the Zeeman energy splitting,
we have a peak instead of the plateau
and $\rho_{xx}$ approaches a universal curve that is 
independent of the magnetic field.
The peak position is scaled by $\sqrt{B}$.
So the presence of the plateau is associated with the Landau level splitting 
between $n=0$ and $n = \pm 1$.
Namely, including the Zeeman energy splitting transforms the peak to the plateau.
For $T<2\Gamma_0$, $\rho_{xx}$ turns to increase again,
and then $\rho_{xx}$ approaches a temperature independent value.
We note that for a conventional parabolic dispersion case $\rho_{xx}$ 
monotonically increases with decreasing the temperature because
the Landau levels are equally spaced.

All features stated above are consistent 
with the experiment \cite{Tajima06} except for $T<2\Gamma_0$.
In the experiment, $\rho_{xx}$ does not approach a temperature independent value
for $T<1$K but increases further with decreasing temperature
changing the slope at a characteristic temperature $T_{\rm exp}$.
This behavior suggests that there is an
another Landau level splitting 
probably associated with valley splitting.
In Ref.~\cite{Kobayashi09}, a Kosterlitz-Thouless transition scenario was proposed.
We will investigate this point further in a future publication.

Now we comment on the tilt of the Dirac cone.
In $\alpha$-(BEDT-TTF)$_2$I$_3$, theoretical calculations
suggest that the Dirac cone is tilted.\cite{Kobayashi07}
In the presence of the tilt of the Dirac cone,
the Landau level wave functions are deformed \cite{Morinari09}
that leads to anisotropy of the resistivity.
However, the features of the inplane magnetoresistance
are unaffected by the tilt.
The temperature dependence of the inplane magnetoresistance
is determined by the Landau level structure.
Since the tilt of the Dirac cone just leads to
a modification of the overall factor
of the Landau level energies and does not affect
the Landau level structure qualitatively,\cite{Morinari09}
the tilt is unimportant for the temperature dependence 
of the inplane magnetoresistance.

\begin{figure}
   \begin{center}
    \includegraphics[width=0.8 \linewidth]{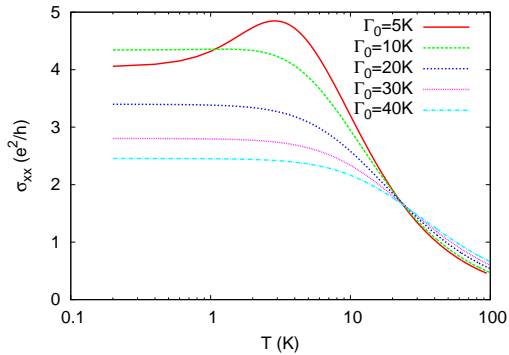}
   \end{center}
   \caption{ \label{fig_graphene_sxx}
	(Color online) The temperature dependence of $\sigma_{xx}$ for graphene
	for different $\Gamma_0$ at $B=10$~T. 
    }
 \end{figure}

Using the theory, 
we are able to understand
some results about $\sigma_{xx}$ in graphene.
Figure \ref{fig_graphene_sxx} shows 
$\sigma_{xx}$ for different $\Gamma_0$ at $B=10$~T.
We computed $\sigma_{xx}$ for $B>10$~T as well (not shown)
and found similar behaviors.
The results with $\Gamma_0>10$~K are 
in good agreement with the experiment \cite{Checkelsky08} for $B<8$~T.
Experimentally $\Gamma_0$ is estimated \cite{Giesbers09} 
as $\Gamma_0 \sim 30$~K.
For clean samples with $\Gamma_0$, we should observe a peak associated with 
the Zeeman splitting around $T = 2\Gamma_0 - \mu_B B$.
For $\Gamma_0=5$~K, the peak appears around $2\Gamma_0 - \mu_B B \sim 3$~K
as shown in Fig.~\ref{fig_graphene_sxx}.
In the experiment reported in Ref.~\cite{Checkelsky08}, 
$\sigma_{xx}$ decreases at low temperatures for $B>10$~T.
To understand this behavior, we need to assume that a valley splitting
occurs as discussed in the literature \cite{Yang07}.

\section{Conclusion}
\label{sec_summary}

In conclusion, we have investigated the inplane resistivity of Dirac fermions
under magnetic field.
We have included the Landau level structure, the Zeeman energy splitting, and
the Coulomb interaction effect between Dirac fermions.
The Coulomb interaction plays an important role at high temperatures
where Dirac fermions are excited from the zero energy Landau level.
We found that the $n=0$ Landau level behaves differently compared
to the other Landau levels.
The features observed in $\alpha$-(BEDT-TTF)$_2$I$_3$ 
are consistent with our result except for $T<1$K
where a valley splitting may play an important role.
This theory has also been applied to graphene. 
We have found a consistent behavior with an existing 
experimental data and have predicted the presence 
of a peak structure of conductivity in clean samples.

\section*{Acknowledgments}
We would like to thank N.~Tajima for helpful discussions.
This work was supported by KAKENHI (21740252),
the Global COE Program "The Next Generation of Physics, 
Spun from Universality and Emergence," 
and Yukawa International Program for Quark-Hadron Sciences at YITP.

\bibliography{../../../references/tm_library2}

\end{document}